\documentstyle{lamuphys}
\makeatletter
\let\chapter\hid@chapter
\makeatother
\input psfig.sty
\def\etal   {{\it et~al.}}

\def\gax    {${_>\atop^{\sim}}$}
\def\lsun {${\rm L_{\bigodot}}$}

\begin{document}
\pagenumbering{arabic}
\title{ISO Observations of Quasars and Quasar Hosts}

\author{Belinda J.\,Wilkes\inst{}}

\institute{Smithsonian Astrophysical Observatory, 60 Garden St.,
Cambridge, MA 02138, USA}

\maketitle

\begin{abstract}
The Infrared Space Observatory (ISO), launched in November 1995,
allows us to measure the far-infrared (far-IR) emission of quasars in 
greater detail and over a wider energy range than previously possible.
In this paper, preliminary results in a study of the 5--200 $\mu m$
continuum of quasars and active galaxies are presented. Comparison of
the spectral energy distributions show that, if the far-IR emission from
quasars is thermal emission from galaxian dust, the host galaxies of
quasars must contain dust in quantities comparable to IR 
luminous galaxies rather than normal spiral
galaxies. In the near-IR, the ISO data confirm an excess due to a warm
`AGN-related' dust component, possibly from the putative molecular torus.
We report detection of the high-redshift quasar, 1202-0727, in the near-IR
indicating that it is unusually IR-bright compared with low-redshift quasars.
\end{abstract}
\section{Introduction}

Quasars are multi-wavelength emitters, emitting roughly equal amounts of
radiation throughout the whole electromagnetic spectrum from
far-IR through to $\gamma$-ray energies. 10\% are also
strong radio emitters. To understand
the energy generation mechanisms at work,
it is first essential to obtain multi-$\lambda$ data covering the full 
spectral range of the emission. We now have a good understanding of the
spectral energy distributions (SEDs) of low-redshift
quasars and active galaxies. However, in the far-IR this has been 
limited by the short lifetime and wide beam of the ground-breaking IRAS
satellite. Now, more than 10 years later, ISO is providing us with 
the chance for a second, more detailed look at the far-IR sky, allowing
us to extend our knowledge to IR-fainter and higher redshift sources
and to longer and shorter wavelengths (5--200$\mu m$).

To this end we are observing a sample of quasars and active galaxies
with the photometer on ISO (ISOPHOT).
The sample was originally designed to include $\sim 130$ quasars and
active galaxies covering the full range of redshift and of known SED
properties. With the reduced in-flight sensitivities of ISOPHOT, our
program has been reduced significantly and will likely include $\sim 50$
objects, not all with full wavelength coverage. The sample will include
full wavelength coverage for a
well-defined subset of optically-selected, PG quasars (\cite{alea}), along
with a few high-redshift quasars, X-ray selected Seyfert 1 galaxies, and red
quasars.

\begin{figure}[h]
\psfig{figure=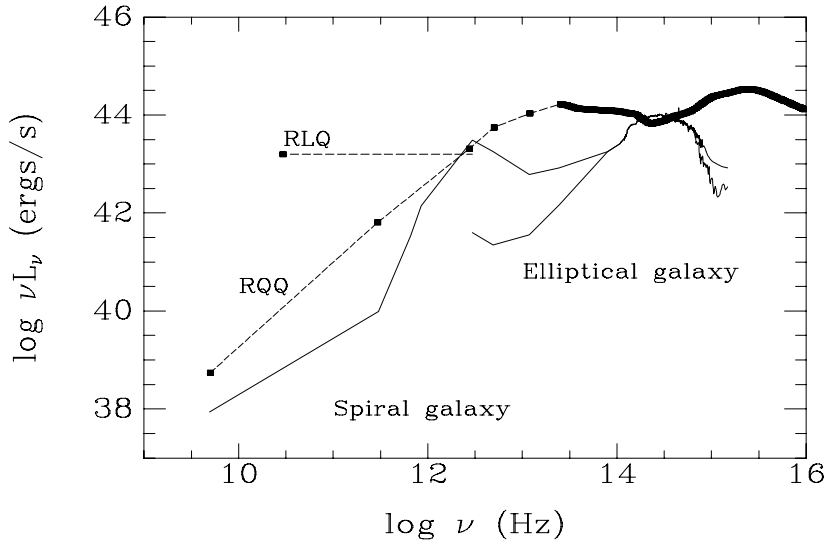,height=3.0truein}
\caption{The Median SED of a low-redshift quasar superposed on
the SED of a spiral galaxy showing the well-explored, near-IR ``window"
($\sim 1 \mu m$) and the potential far-IR ``window" on a quasar's host galaxy
(courtesy Kim McLeod).}
\vskip -0.5in
\label{galsed}
\end{figure}

One question that the ISO data will address is particularly relevant to this
conference, namely the contribution of the host galaxy in 
the far-IR. Figure~\ref{galsed} shows SEDs of spiral and elliptical
galaxies superposed
on the SED of a median low-redshift quasar (from \cite{QED1}). The plot
clearly shows the near-IR ($\sim 1-2 \mu m$ ``window" on the host galaxy
which has been used to great advantage (\cite{mr95,Dunlop93}). 
Although the strength of the far-IR peak, due to cool
dust, is as yet unknown, Figure~\ref{galsed} demonstrates that this is the
most likely wavelength range for a second ``window" on the host galaxy.

\section{ISO Observing Program}
ISOPHOT observations are being made in eight
broad bands covering the full energy range of the instrument, 5--200
$\mu m$. The detector/filter combinations are: P1:5,7,12 $\mu m$;
P2: 25 $\mu m$; C100: 60,100 $\mu m$ and C200: 135,200 $\mu m$.
Until October 1996, AOTs P03 and P22 were used in rectangular chopping
mode with a chopper throw of 180$"$ in all cases. The apertures,
chosen to match the instrument field of view or
the point spread function as applicable, are 52$"$ for
$\lambda \leq 12 \mu m$, 120$"$ for $\lambda = 25 \mu m$ and the array
size for the long wavelength points. For the
largest and/or brightest sources, we
use staring mode with a separate sky observation.
Following the recommendations of the ISOPHOT team, we have re-specified
our remaining time to observe a smaller subset of objects,
re-observing where necessary, using small rasters whose dimensions depend
on the detector in use. These observations are scheduled to begin in early
1997 and should provide the reliable long-wavelength data which is
currently lacking.

\section{Analysis of ISOPHOT data.}

ISOPHOT suffers from several well-known problems which complicate the
data analysis and limit (currently) the accuracy with which
fluxes can be determined (for details see \cite{photman} and associated
updates):
\begin{itemize}
\item The responsivity of all detectors drifts significantly
following a change in the incident signal, for example when
pointing to a new source or changing filters. This drift is difficult to
calibrate and the analysis software does not yet support fitting for 
chopped observations such as ours.
We have thus concentrated our efforts on analysing objects with
observations sufficiently long that the detector reaches a stable portion of
the drift curve, generally \gax 128 secs total time. 
\item The internal (FCS) calibrators needed to be re-calibrated in-orbit
following a change of state of FCS1 in February 1996. The combination of
the shortness of the FCS observations (16/32 secs) taken during an
individual observation 
and the lack of a definitive re-calibration led us to use 
the default responses for all detectors in this
analysis. There is a drift $\sim 30$\% in the detector responsivity
as a function of time during an orbit so our flux normalisation
errors are expected to be of this order.
\item At long wavelengths (C200 detector in particular) the two adjacent
beams are large enough that
part of the detector lies within a part of the telescope beam which is
significantly vignetted. This leads to an asymmetry in the derived fluxes
across the detector and a $\sim 20$\% flux correction. The correction
for this affect has not yet been released and has not been applied
to our data.
\end{itemize}

Our analysis was performed using the PHOT Interactive Analysis Package (PIA),
an IDL-based system provided by the PHOT team. We carried out the following
steps: non-linearity correction, read-out de-glitching, 1st order 
fit to ramps,
dark current subtraction, de-glitching to delete highly discrepant 
points, deletion of data during strong detector
drifts and of remaining highly-discrepant points, background subtraction, and
calibration using the default responsivities. Background subtraction
is done using the average of the background in the chopper plateaux before 
and after
each source plateau. On a non-linear drifting curve, this is not accurate
and adds noise to the signal which could be reduced by fitting the
background and source curves separately and then subtracting them. Currently 
points on the drift curve are deleted, reducing the potential S/N of the
observations once more sophisticated analysis is carried out.

From the subset of our sources with relatively long exposure times, we chose
four objects covering a range of luminosity and redshift: PG1244+026, a
radio-quiet Seyfert 1 galaxy; PG1543+489, a radio-quiet quasar;
3C249.1 (PG1100+772), a radio-loud quasar; and 1202-0727, a high-redshift,
radio-quiet quasar. 
The observational details are provided in Table~\ref{obstab}.
For these sources we found reliable detections in most/all the short
wavelength bands (5-25 $\mu m$). In the long wavelength bands, however,
only two of the sources are detected, the other two have negative ``detections"
in all four long wavelength bands. 

Subsequent analysis of additional sources not reported here
has shown a large number of negative signals at long
wavelengths, particularly 135,200 $\mu m$. We are currently investigating the
cause and, since the negative signals are often at levels similar to the
positive ones, are treating all our long wavelength data with
scepticism. The most likely cause is cirrus confusion and would imply
significant structure on the scale of $\sim 3'$ (our chop 
distance). However further investigation is required to confirm this.

\begin{table}
\caption{Details of the ISOPHOT observations.}\label{obstab}
\begin{tabular}{llllllllllll}
\hline
Name & z& ISO date & AOT$^1$ & Filter & Time & Filter & Time & 
Filter & Time & Filter & Time \\
&&&& $\mu m$ & sec$^2$ & $\mu m$ & sec$^2$ & $\mu m$ & sec$^2$ & $\mu m$ &
sec$^2$ \\
\hline
PG1244+026 & 0.048 & 14/07/96 & PHT03 & 4.85 &16& 7.3 &256& 12 &256& 25 &256\\
&&&PHT22 & 60 & 64 & 100 & 64 & 135 & 64 & 200 & 64 \\
PG1543+099 &0.400 & 30/05/96 & PHT03 & 4.85 & 256 & 7.3 & 128 & 12 & 128 &
25 & 256 \\
&&& PHT22 & 60 & 64 & 100 & 64 & 135 & 64 & 200 & 128 \\
3C249.1 &0.313& 17/06/96 & PHT03 & 4.85 & 256 & 7.3 & 256 & 12 & 256 &
25 & 256 \\
(PG1100+772) &&& PHT22 & 60 & 128 & 100 & 128 & 135 & 256 & 200 & 256 \\
1202$-$0727 &4.69& 19/07/96 & PHT03 & 4.85 & 512 & 7.3 & 512 & 12 & 512 &
25 & 512 \\
&&& PHT33 & 60 & 512 & 100 & 128 & 135 & 512 & 200 & 512 \\
\hline
\end{tabular}
\begin{minipage}{6.5in}
1: AOT: Astronomical Observation Templates \\
2: on-source time 
\end{minipage}
\end{table}

\section{Spectral Energy Distributions}
\begin{figure}
\psfig{file=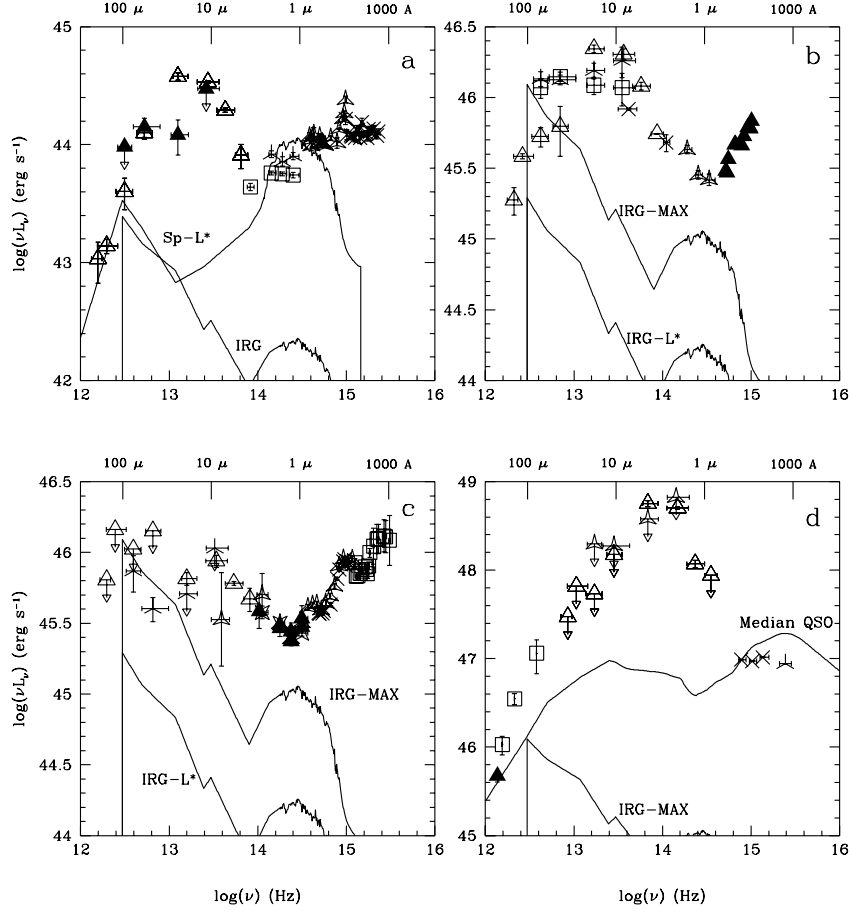,height=5.3in}
\caption{The far-IR -- ultra-violet SEDs of a: PG1244+026, b:
PG1543+489, c: 3C249.1, d: 1202-0727, with various galaxy and quasar SEDs
(as labelled) superposed. Each different dataset uses a different symbol,
the ISO points are always indicated by open triangles.}
\label{sedfig}
\end{figure}

We have combined our ISO results with data from other wavelengths collected
by ourselves and from the literature to generate SEDs
of the four objects (Figure~\ref{sedfig}). 
To investigate possible contributions from the quasar host galaxy, 
particularly in the far-IR, SEDs for several kinds of galaxies 
were generated using data from the literature (\cite{kkmpc}).
The IRG and ULIRG templates correspond roughly to L$_{IR} \sim 10^{10-11}$
\lsun and $10^{11.5}$ \lsun respectively.
Superposed on each quasar SED, we have shown various galaxy SEDs as labelled
and a median SED for low-redshift quasars (\cite{QED1}) for direct
comparison (Figure~\ref{sedfig}). 

PG1244+026 is a low luminosity active galaxy (L$\sim 10^{44}$ erg s$^{-1}$),
officially classified as a Seyfert 1 galaxy and 
with low redshift (z=0.048). Figure~\ref{sedfig} shows an 
L$^{*}$ spiral galaxy which is consistent with the AGN SED $\sim 1 \mu m$
and with the cool dust contribution in the far-IR. Between
5 and 100 $\mu m$ the quasar SED is dominated by an ``AGN" component
believed to originate in warm dust within the putative molecular torus. 
This component peaks $\sim 25-60 \mu m$ in PG1244+026.

PG1543+489, also an optically selected PG quasar, has a luminosity$\sim
10^{45.5}$ erg s$^{-1}$
(a bona fide quasar) and a redshift of 0.400. 
In this case an L$^{*}$ galaxy would make no significant constribution
in the near-IR. A galaxy with the maximum host galaxy luminosity seen 
to date (\cite{mr95}) is a factor $\sim 4$ too low at 1$\mu m$.
An amount of dust comparable to an IR-bright galaxy is necessary to 
explain the far-IR emission (L$_{IR}$ $\sim 10^{10-11}$ \lsun).
Once again a mid-IR bump due to warm
dust is apparent, peaking $\sim 100 \mu m$.

\begin{figure}[t]
\psfig{file=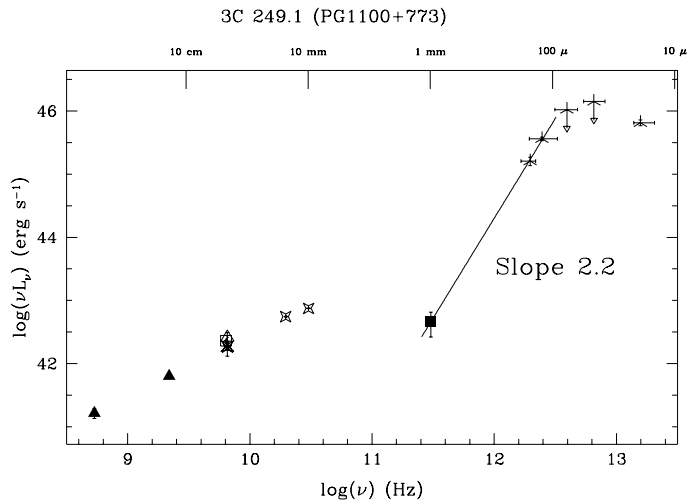,height=4.0in}
\vskip -1.2in
\caption{The radio--far-IR SED of 3C249.1 showing the current limits on
the slope of the far-IR turnover.}
\label{3c2491}
\end{figure}

3C249.1 is a lobe-dominated, radio-loud quasar at a reshift 0.389.
The ISO short wavelength detections once again show a typical
mid-IR bump with a peak $< 100 \mu m$. The IRAS upper
limits (\cite{QED1}) are very low ($\sim$ 20 mJy at 100 $\mu m$)
and inconsistent with the ISO detections. Re-analysis of the IRAS data
using the current software (\cite{xscanpi})
yields upper limits which are largely consistent 
(Figure~\ref{sedfig}c) so we conclude that the earlier numbers are in error.
Unfortunately we currently have only weak upper limits on the far-IR emission
of this source indicating that the data could be consistent with an L$^{*}$,
IR-bright galaxy but not one with maximal luminosity in the near-IR.  
However since we have no estimates of the host galaxy from the near-IR,
this provides a very weak constraint on the amount of dust.
Figure~\ref{3c2491} shows the far-IR SED with a detection at 1mm. 
Assuming that the 1mm flux represents the long wavelength tail of the
far-IR emission, our current data give an upper limit to the slope
of the far-IR turnover of $\alpha < 2.2$ (f$_{\nu} \propto \nu^{\alpha}$).
The discontinuity between the far-IR and radio emission in the SEDs of
lobe-dominated, radio-loud quasars is generally interpreted as
evidence for differing emission mechanisms and thus thermal IR 
emission (\cite{aba89}).
However, the flat far-IR turnover in this source prevents us from
ruling out non-thermal synchrotron emission.

1202-0727 is one of the highest redshift quasars known (z=4.690).
It is an extremely interesting source, mentioned a number of times
in these proceedings (Barvainis, Yamada). 
Optically it is a double source with $4"$ separation, has strong CO
emission (\cite{Ohta96,Omont96}), a sub-mm spectrum which suggests 
emission from 50--100 K dust (\cite{Iea94}) and a Ly$\alpha$ emission 
companion 2$"$ away with the same redshift (\cite{hu96}). 
The source has a very high luminosity, $\sim 10^{47}$ erg s$^{-1}$,
such that the contribution from its host galaxy in the rest-frame
optical and near-IR is several orders of magnitude 
below that seen by ISO (12, 25 $\mu m$ observed frame, Figure~\ref{sedfig}d).
The mid-IR bump, with a broad peak from 4--80 $\mu m$ in the rest frame,
is two orders of magnitude stronger than that of the low-redshift median 
(Fig.~\ref{sedfig}d).  The rest-frame far-IR emission
determined by the sub-mm observed frame data (\cite{Iea94})
would require a host galaxy similar to an ultra-luminous IR galaxy (ULIRG)
in a pure dust scenario. The crude far-IR upper limits from ISO suggest
that our planned re-observation could strongly constrain the mid-IR SED.

\section{Conclusions}
Although the current status of the ISO far-IR data limits the usefulness of
ISO to study the host galaxy dust contribution, we can already demonstrate
that, if the far-IR emission of bona fide quasars (L$> 10^{44}$ ergs$^{-1}$)
is from the host galaxy, these galaxies are unusually far-IR bright,
comparable to IR-bright galaxies or ULIRGs (L$_{IR} \sim 10^{10-11.5}$ \lsun).
The ISO data also allow us to 
investigate the mid-IR ``AGN" bump in quasars covering a range of redshift 
and luminosity. We plan to use these data to test and constrain current 
models of emission from a molecular torus (\cite{pk92,rr95}).

We have detected the z=4.69 quasar, 1202-0727, in the rest-frame near-IR
at a level far above that seen in typical low-redshift quasars. 
Observations of more high-redshift quasars are necessary to determine whether
the near-IR emission is unusual, as are many other aspects of this source.
For pure host galaxy far-IR emission, the host must be comparable to
an ULIRG to explain the mm data (\cite{Iea94}).

\section*{Acknowledgements}
I would like to thank all my collaborators on this project, in
particular Drs. Kim McLeod, Jonathan McDowell and Martin Elvis at CfA
and our other ISO co-Is. Thanks are also due to the ISOPHOT team in Heidelberg
and the ISO centers at VILSPA and IPAC for their prompt and invaluable help
in response to my
frequent email messages. The financial support of NASA grant NAGW-3134 is
gratefully acknowledged.

%

%
%


\begin{thebibliography}
%
%
\bibitem{}{aba89}{Antonucci \etal~1990} Antonucci,~R.,
Barvainis,~R. \& Alloin,~D. 1990, ApJ, 353, 416
\bibitem{}{Dunlop93}{Dunlop \etal\ 1993} Dunlop,~J.~S.,
Taylor,~G.~L., Hughes,~D.~H. \& Robson,~E.~I. 1993, MNRAS
264, 455
\bibitem{}{rr95}{Efstathiou \& Rowan-Robinson 1995} Efstathiou,~A. \&
Rowan-Robinson,~M. \& 1995, MNRAS 273, 649
\bibitem{}{QED1}{Elvis \etal~1994} Elvis,~M.,
Wilkes,~B.~J., McDowell,~J.~C., Green,~R.~F.,
Bechtold,~J., Willner,~S.~P., Cutri,~R., Oey,~M,~S., and Polomski,~E.
~1994, ApJS, 95, 1
\bibitem{}{hu96}{Hu \etal\ 1996} Hu,~E.~M., McMahon,~R.~G. \& Egami,~E.
1996, ApJ, 459, L53
\bibitem{}{Iea94}{Isaak \etal\ 1994}
Isaak,~K.~G., McMahon,~R.~G., Hills,~R.~E. \& Withington,~S. 1994 MNRAS,
269, L28
\bibitem{}{photman}{ISOPHOT Observers Manual} ``ISOPHOT Observers Manual"
and associated updates found on WWW:
http://isowww.estec.esa.nl:80/ISO/iso\_manuals.html
\bibitem{}{alea}{Laor \etal~1996}
Laor,\, A., Fiore,\, F., Elvis,\, M., Wilkes,\, B.J., \& McDowell,\,
J.C. (1996) ApJ. {\it in press}
\bibitem{}{mr95}{McLeod \& Rieke 1994}
McLeod, K.K. \& Rieke, G. 1995, ApJ, 441, 96
\bibitem{}{kkmpc}{McLeod, private communication} McLeod, K.K. private
communcation
\bibitem{}{Ohta96}{Ohta \etal~1996} Ohta,~K,
Yamada,~T., Nakanishi,~K., Kohna,~K.,
Akiyama,~M. \& Kawabe,~R, 1996, Nature, 382, 4260
\bibitem{}{Omont96}{Omont \etal~1996} Omont,~A.,
Petitjean,~P., Guilloteau,~S.,
McMahon,~R.~G., Solomon,~P.~M. \& Pecontal,~E. 1996, Nature, 382, 4280
\bibitem{}{pk92}{Pier \& Krolik 1992} Pier,~E.~A. \& Krolik,~J.~H.
1992, ApJ 401, 99 
\bibitem{}{xscanpi}{XSCANPI} XSCANPI, Interactive software for the
analysis of IRAS data
provided by IPAC via ``telnet xscanpi.ipac.caltech.edu"
%
\end{thebibliography}
\end{document}